\documentclass[preprint,review,12pt]{elsarticle}
\usepackage{longtable}
\usepackage{amssymb}
\usepackage{epsfig}
\journal{New Astronomy}

\begin{document}

\begin{frontmatter}

\title{KHz QPOs in LMXBs, relations between different frequencies and compactness of stars.}

\author{Taparati Gangopadhyay $^{1}$, Xiang-Dong Li $^{2}$, Subharthi Ray $^{3}$,\\
 Mira Dey $^{\dagger{1}}$, Jishnu Dey $^{\dagger{1}}$ }

%\maketitle
 
\address{$^1$ Dept. of Physics, Presidency University, Kolkata 700 073, India.\\
$^{2}$ Department of Astronomy, Nanjing University, Nanjing, P. R. China 210093 \\ 
$^3$ Astrophysics and Cosmology Research Unit, School of Mathematical Sciences, Univ. of KwaZulu-Natal, Durban 4000, South Africa.\\
$^\dagger$ Associate, IUCAA, Pune; Emeritus Research Scientist, Supported by Department of Science and Technology (Govt. of India). 
}

%\maketitle

\begin{abstract}
We suggest that the mass of four compact stars SAX J1808.4$-$3658, KS 1731$-$260, SAX J1750.8$-$2900 and IGR J17191$-$2821 can be determined from the difference in the observed kiloHertz quasi periodic oscillations (kHz QPO-s) of these stars. The stellar radius is very close to the marginally stable orbit $R_{ms}$ as predicted by Einstein's general relativity. It may be noted that the first of these stars was suggested to be a strange star more than a decade back by Li \emph{et al.} (1999) from the unique millisecond X-ray pulsations with an accurate determination of its rotation period. It showed kHz QPO-s eight years back and so far it is the only set that has been observed. This is the first time we give an estimate of the mass of the star and of three other compact stars in Low-Mass X-ray Binaries using their observed kHz QPO-s. 
\end{abstract}

%\pacs{98.70.Vc,95.75.Pq,98.80Es}
%\maketitle
\end{frontmatter}

\section{Introduction}

Strong variability on millisecond time scales, the kilohertz quasi-periodic oscillations (kHz-QPOs) observed in the low mass X-ray binaries (LMXBs) was first reported by van der Klis \emph{et al.} (1996), with observations from the Rossi X-ray Timing Explorer (RXTE). Ever since, it led to many theoretical modelling and prediction of its origin, and to determine the \emph{correct} equation of state (EOS) of the compact objects (see van der Klis, 2006 for a recent review). Although in its earlier stage of detection it was found in the Atoll and Z class of sources of the LMXBs, later it was also observed in the millisecond X-ray pulsars and a few uncategorised sources.

Typically these characteristic frequencies have two peaks, the lower ($\nu_{low}$) one ranges from a few hundreds of hertz and the upper ($\nu_{up}$) one often goes beyond a kilohertz. Although some QPOs with both the frequencies falling much below the kilohertz range have been observed in some black hole candidates, for the present study we shall refrain from considering them.

\begin{figure}
\centering
\psfig{file=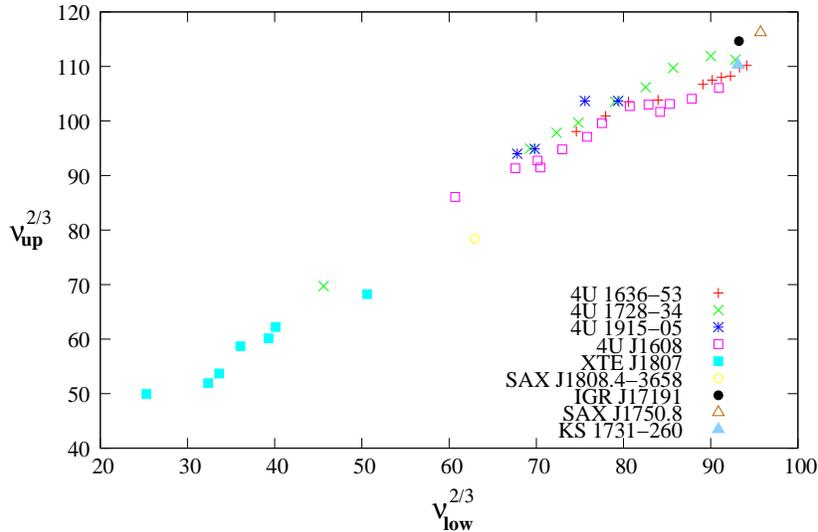,height=7cm}
\caption{Plot of $\nu_{up}^{2/3}$ vs. $\nu_{low}^{2/3}$ for different LMXBs. Different symbols and colours have been used for different sources.}
\label{fig:nuup_nulow}
\end{figure}

Here we start from a recent argument if the kHz-QPO-s are related in any sort of way to the spin frequency of the star. M\'endez and Belloni (2007) re-examined all reliable available data of sources for which there exist measurements of two simultaneous kHz QPO-s and spin frequencies ($\nu_s$) and advance the possibility that the difference between the QPO-s, $\Delta \nu$ (the difference between the upper and the lower frequencies) are not related to each other. We have used their data for these ten sources and we find a correlation between the two simultaneous kHz QPO-s raised to the power 2/3 (Fig \ref{fig:nuup_nulow}). Since QPO-s are represented by peaks in the power spectrum, they are related to the relatively stable orbits of accreting matter from which particles fall onto the compact object - in our case assumed to be a strange star - and these orbits are the maximally stable orbit given by an average radius of $R_{ms}$ for $\nu_{up}$ and the outermost orbit $R_0$ for $\nu_{low}$. This assumption leads to possible determination of the mass and radius of the compact strange star.

%\newpage

Turning to history, the most popular model to explain the difference of kHz QPOs was the beat frequency model (Strohmayer \emph{et al.}, 1996) which assumed the upper QPO ($\nu_{up}$) as the Keplerian orbital frequency $\nu_{k}(r)$  of the innermost orbit in the accretion disk around the star, the separation between the two peaks $i.e.$ $\Delta \nu_{peak}=\nu_{up}-\nu_{low}$ as $\nu_{spin}$ and the lower QPO  ($\nu_{low}$) as the beat frequency of $\nu_{k}(r)$ and $\nu_{spin}$. This model had suggested that for a particular star
$\Delta \nu~=~\nu_{spin}$ will remain constant. However, it has been observed that in many sources the $\Delta \nu$ is not related to $\nu_{spin}$.Also $\Delta \nu$ changes with time. The QPO resonance model of Kluzniak and Abramowicz (2001) is a beautiful model but comes with its own set of problems. It states that the QPO could be explained by non linear resonant motions of accreting fluid in strong gravity. The two QPO frequencies depend here on the amplitude of the oscillatory motion and are approximately in the ratio 3:2. However, the difficulties mainly arise from the fact that QPO frequencies are not consistently equal to this mentioned ratio, and the twin peaks shift their positions (Rebusco, 2008). Although all of the present QPO models come with some conceptual difficulty or the other, for our present work, we shall adhere to the QPO model of Titarchuk and Osherovich (1999) which allowed us in an earlier paper to determine the mass and limits
to the radius of the accreting star (Li \emph{et al.} 1999).

Most compact stars showing a pair of QPO-s in the X-ray power spectrum, have variable $\delta\nu$. The first star to show accurately constant X-ray pulse period was the X-ray pulsar SAX J1808.4$-$3658.  This source was found to show kHz QPO-s with the a frequency difference of 195 kHz. This is roughly half the spin frequency of the star $\nu_{spin}$ which is measured very accurately to be 401 Hz. So a modified beat frequency model was tried with $\nu_{peak} \sim 0.5 \nu_{spin}$. But for many other stars the difference of kHz QPO-s has no apparent correlation with the spin frequency of the star like that has been suggested by the beat frequency model. 

There are now at least five other X-ray pulsars whose spin is steady and accurately measured and only for one of these, XTE J1807$-$294 kHz QPO-s have been seen. We find, from applying the Titarchuck and Osherovich(1999)
model, that the masses (M) and radii (R) of these two stars are very small and may fit those of
strange stars.

\section{Details of Calculation and Results}
\subsection{Derivation of the mass and a limit to the radius of the compact star, using QPO frequencies}
We assume the possibility  that the lower QPO frequency is due to particles thickly clustered at the inner edge of the accretion disc (at a radial distance $R_0$) and the upper QPO frequency is due to particles thickly clustered at the marginally stable orbit (at a radial distance $R_{ms}$). The expression for the marginally stable radius (Bardeen, Press and Teukolsky, 1972) is given as: 

\begin{equation} 
R_{ms} = R_g\left\lbrace 3+Z_2-[(3-Z_1)(3+Z_1+2Z_2)]^{1/2}\right\rbrace 
\end{equation} 
where 
\begin{equation}
Z_1 = 1+[1-(a/R_g)^2]^{1/3}[(1+a/R_g)^{1/3} +(1-a/R_g)^{1/3}]
\end{equation}
\begin{equation}
Z_2 = [3(a/R_g)^2 +Z_1^2]^{1/3} 
\end{equation}
where $R_s=2GM/c^2$, $R_g=GM/c^2$ and $a= 2 \pi \nu_{spin} I/Mc$, $I$ is the moment of inertia, $c$ is the velocity of light and $G$ is the gravitational constant. A lower limit for $R_{0}$ according to the Titarchuck and Osherovich model is as follows (Li \emph{et al.}, 1999b)
\begin{equation} 
R_0 = 3 R_{s}\times \left(\frac{M}{M_{\odot}}\right)^{1/3}
\end{equation}
and hence it is clear that $R_{0}$ depends only on $M$ and whereas $R_{ms}$ depends on both $M$ and $\nu_{spin}$. So for a particular star (fixed $M$ and $\nu_{spin}$) $R_{ms}$ should be constant. From the Fig(\ref{fig:nuup_nulow}) above we see that  ${\nu_{up}^{2/3}}/{\nu_{low}^{2/3}} ~\sim~ 1.2 - 1.3$ for most of the stars and one can calculate the most probable outermost orbit $r_0$ and consider its relation to the $R_0$ given above. We tabulate $\nu_{low}$, $\nu_{up}$ and $\nu_{spin}$ for 10 sources in Table \ref{list} for which Fig(\ref{fig:nuup_nulow}) shows that ${\nu_{up}^{2/3}}/{\nu_{low}^{2/3}}$ is nearly constant.  

\begin{center}
\begin{longtable}{|c|c|c|c|}
\caption{kHz QPO frequencies and spin frequencies for different stars} \label{list} \\
%\begin{tabular}{lccc}
\hline 
Sources & $\nu_{low}$ & $\nu_{up}$ & $\nu_{spin}~ in ~Hz$\\
\hline \multicolumn{4}{|c|}{\bf X-Ray pulsars} \\
\hline
SAX J1808.4$-$3658   & 499 &  694 & 401\\
\hline
XTE J1807$-$294    & 127.1 &  352.8 & 191.0 \\
             &  184  &  374   &     \\
             &  195  &  393.4 &      \\
             & 216.5 &  449.6 &      \\
             & 246.3 &  466.3 &      \\
             & 253.9 &  490.6 &      \\
             & 359.9 &  563.6 &      \\
 
\hline \multicolumn{4}{|c|}{\bf other stars}\\
\hline
KS 1731$-$260      &  898.3  & 1158.6 & 524  \\
\hline
IGR J17191$-$2821  & 900   & 1227 & 294\\
\hline
SAX J1750.8$-$2900 & 936   &  1253  & 600.75 \\  
\hline
4U 1728$-$34       &  576.6 & 925.0 &  363 \\
                   &  614.9 &  967.6   &     \\
                   &  647.1 &  995.2 &      \\
                   &  702.3  &  1053  &     \\
                   &  749.7 &  1094.2&      \\
                   &  793.0  &  1148.9   &     \\
                   &  853.8 &  1183.2 &      \\
                   &  894.2 &  1172.9  &     \\
                   &  308  &  582    &      \\ 
\hline
4U 1608$-$52& 472.8 &  798.3 & 619.0 \\
             &  555.6  &  872.5   &     \\
             &  561.3  &  874.5 &      \\
             & 587.3 &893.0   &      \\
             & 623.2 &922.7   &      \\
             & 660.0 &956.5   &      \\
             & 682.5 &993.7   &      \\
             & 725.0 &1040.9   &      \\
             & 754.0 &1044.9   &      \\
             & 772.3 &1024.9   &      \\
             & 787.7 &1047.0   &      \\
             & 822.7 &1061.5   &      \\
             & 866.9 &1092.2   &      \\
\hline
4U 1636$-$53         &  644    &  971 & 582 \\
                     &  688  & 1013   &     \\
                     &  723  &  1053 &      \\
                     &  769  &  1058  &     \\
                     &  841  &   1102&      \\
                     &  856  &  1114  &     \\
                     &  871 &  1122&      \\
                     &  886 &  1126   &     \\
                     &  901  &  1150 &      \\
                     &  913  &  1156  &     \\
\hline
4U 1702$-$43       &  657.1  & 1000.1 & 330  \\
                   &  707.7  &  1037.7  &     \\
                   &  770.0 &  1084.8  &     \\

\hline
4U 1915$-$05      & 706.9  & 1055.3  & 272  \\
                   &  656.9 &  906.4 &     \\
                   &  558.3  &  911.2 &      \\
                   &  583.2  &  923.7 &   \\ 
\hline 
\end{longtable}
\end{center}

From equations (1) to (4) we obtain the difference 
\begin{equation}
\nu_{up} - \nu_{low} = [(R_0/R_{ms})^{3/2} - 1] \nu_{low},
\end{equation}
so that we can seek the stellar mass, the corresponding stellar radius and the moment of inertia to fit this difference. If the difference remains constant as in the case of SAX J1808.4$-$3658 (at 195 Hz) we can predict the mass to be 0.904 $M_{\odot}$. For this mass and spin the $R_{ms} = 7.409$ km. For strange stars, the corresponding radius is known from our EOS (see section 2.2) to be 6.89 km which is nicely inside $R_{ms}$. The derived value of $r_0 = 8.607$ km and the approximate relation $R_0 ~=~ 3 \times 2.9532 M^{1/3}$ gives a result $R_0 = 8.566$ km which is very close to $r_0$ such that our model leads to a consistent result. A recent estimate of the mass of SAX J1808.4$-$3658 by Elebert \emph{et al.}, 2009, from its optical spectroscopy and photometry showed that its mass is $0.9\pm 0.4 ~M_\odot$. This is consistent with the mass
derived according to the strange-star model.

\subsection{The mass-radius relation of strange stars}
For use in this paper, we have fitted the mass $M/M_{\odot}$ of the strange star model (Dey \emph{et al.}, 1998) along with the corresponding moment of inertia $I$, as a polynomial of the radius R, for the range where the radius increases monotonically with mass as follows:

\begin{equation}
M/M_{\odot} = a1 R^{b1} + a2 R^{b2}
\end{equation}  

\begin{equation}
I  = g1 R^{h1} + g2 R^{h2}
\end{equation}

The parameters are given in Table \ref{tab:param}. This is for the EOS A which gives the right masses and magnetic moments of baryons in a finite system calculation with our chosen interquark force as given in Bagchi \emph{et al.} (2006). We would like to stress that this SS model has the Richardson potential modified with two appropriate but different scales for confinement and asymptotic freedom for the interquark force. The parameters of this force are all fixed from elaborate baryon mass and magnetic moment calculations and satisfies the stringent binding energy requirement that 
\begin{equation}
E/A(uds)~ \le ~E/A(Fe^{56})~\le~E/A(ud).
\end{equation}

\begin{table}
\begin{center}
\caption{Parameters for fitting mass and moment of inertia for EOS A}
\label{tab:param}
\vskip .2cm
\begin{tabular}{|c|c||c|c|}
\hline
Parameters & $M/M_{\odot}$ & Parameters & I \\
\hline
a1 & $4.6\times 10^{-2}$ & g1& $8.9\times 10^{-4}$ \\ 
a2 & $3.0\times 10^{-6}$ & g2& $1.7\times 10^{-7}$ \\
b1 & 3.0 & h1 & 5.0 \\
b2 & 7.2 & h2 & 9.0\\
\hline
\end{tabular}
\end{center}
\end{table}

\subsection{Application to other Low-Mass X-ray Binaries}
For KS 1731$-$260, SAX J1750.8$-$2900 and IGR J17191$-$2821, the masses and radii are 0.878, 0.895, 0.843 in $M_{\odot}$ and 6.84, 6.88, 6.762, $R_{ms}$ = 7.00, 7.031, 7.051  in km respectively. One can see the closeness of $R_{ms}$ to the stellar radius and this was what had made the estimate for 4U 1728$-$34 reliable in the paper by Li \emph{et al.}, (1999b).

For the other millisecond pulsar XTE J1807$-$294 if we may estimate the mass and the radius to be  $M ~=~ $0.782 $M_{\odot}$ and $R ~=~ 6.62$ km with the $R_{ms}$ very close $\sim$ 6.672 km and $R_0$ to be 8.162 km - then for the seven sets of QPO-s given by M\'endez and Belloni (2007), we find the $r_0$ to be 13.177, 10.705, 10.652, 10.859, 10.210, 10.350 and 8.997 km. Since $R_0$ is just a limit - our results could mean that the outer orbit varies for this and the other stars for example, due to variations in the accretion rate.

\section{Discussions}

For the last few years, various people have searched for a relationship between QPO frequencies and spin frequency of the star. As an example, Belloni, M\'endez and Homan (2005, 2007) have shown that $\nu_{low}$ and $\nu_{up}$ are strongly correlated along a curve passing through the origin. Again, Yin \emph{et al.} (2007) showed that, for six Atoll sources, the relation between $\Delta\nu$ and $\nu_{spin}$ can be fitted by a linear relation 
\begin{equation}
\left\langle \Delta\nu\right\rangle  = -(0.19 \pm 0.05)\nu_{spin} + (389.40 \pm 21.67)~\rm{Hz}
\end{equation}
However, according to these authors, this anti-correlation is just a conjecture since it is based on data of only six sources. As we have stated before, examining all available data of sources for which there exist measurements of two simultaneous kHz QPO-s and $\nu_s$, M\'endez and Belloni (2007) have advanced the possibility that $\Delta\nu$ and $\nu_s$ are not related to each other. On the other hand, we find a correlation between $\nu_{up}^{2/3}$ and $\nu_{low}^{2/3}$ which is expected from the Titarchuck and Osherovich model (our preferred model) in agreement with the findings of Belloni, M\'endez and Homan (2007). Though presently we have data for ten sources only, we hope with future advanced X-ray satellites, like the ASTROSAT satellite to be
launched by India in 2011, we shall be able to get more data on kHz QPO-s which will help us to re-examine findings of the present paper.

\section*{Acknowledgement}
SR acknowledges the NRF SA-India bilateral and the UKZN Competitive Research support. The authors are grateful to Mariano M\'endez for providing the related data and also for confirming that there is only one set of kHz QPO-s observed for SAX J1808.4$-$3658.

%%%%%%%%%%%%%%%%%%%%%%%%%%%%%%%%%%%%%%%%%%%%%%%%%%%%%%%%%%%%%%%%%%%%%%%%%
%%%                REFERENCES
%%%%%%%%%%%%%%%%%%%%%%%%%%%%%%%%%%%%%%%%%%%%%%%%%%%%%%%%%%%%%%%%%%%%%%%%%

\end{document}